\documentclass[lnicst,sechang,a4paper]{llncs}

\usepackage{amssymb}
\usepackage{subcaption}
\usepackage{graphicx}
\usepackage{amsmath}
 \usepackage{amsfonts}
 \usepackage{epsfig}
 \usepackage{multirow}
 \usepackage{wrapfig}

 \usepackage{grffile}

\usepackage{url}
\urldef{\mailsa}\path|{stefan.goeller, alfred.hofmann, peter.strasser, lnicst}@springer.com|
\usepackage[pdfpagelabels,hypertexnames=false,breaklinks=true,bookmarksopen=true,bookmarksopenlevel=2]{hyperref}

\usepackage{amssymb}
\usepackage{pifont}

\begin{document}

\mainmatter  

\title{On the Feasibility of Malware Authorship Attribution}

\titlerunning{Lecture Notes of ICST: Authors' Instructions}

%
%

\author{Saed Alrabaee%
\and Paria Shirani \and Mourad Debbabi \and Lingyu Wang}  %
\authorrunning{Authorship Attribution}



\institute{Concordia University, Montreal, Canada}



%
%

\toctitle{Lecture Notes in Business Information Processing}
\tocauthor{Authors' Instructions}
\maketitle

\begin{abstract}

There are many occasions in which the security community is
interested to discover the authorship of malware binaries, either
for digital forensics analysis of malware corpora or for thwarting
live threats of malware invasion. Such a discovery of authorship
might be possible due to stylistic features inherent to software
codes written by human programmers. Existing studies of authorship
attribution of general purpose software mainly focus on source code,
which is typically based on the style of programs and environment.
However, those features critically depend on the availability of the
program source code, which is usually not the case when dealing with
malware binaries. Such program binaries often do not retain many
semantic or stylistic features due to the compilation process.
Therefore, authorship attribution in the domain of malware binaries
based on features and styles that will survive the compilation
process is challenging. This paper provides the state of the art in
this literature. Further, we analyze the features involved in those
techniques. By using a case study, we identify features that can
survive the compilation process. Finally, we analyze existing works
on binary authorship attribution and study their applicability to
real malware binaries.

\end{abstract}

\section{Introduction} 
Authorship attribution comprises an important aspect of many
forensic investigations, which is equally true in the computer
world. When a malware attacks computer systems and leaves behind a
malware corpus, an important question to ask is '\emph{Who wrote
this malware?}'. By narrowing down the authorship of a malware,
important insights may be gained to indicate the origin of the
malware, to correlate the malware to previously known threats, or to
assist in developing techniques for thwarting future similar
malware. Considering the fact that humans are creatures of habit and
habits tend to persist, therefore, various patterns may be embedded
into malware when their creators follow their habitual styles of
coding.

Although significant efforts have been made to develop automated
approaches for source
code~\cite{caliskan2015anonymizing,krsul1997authorship,spafford1993software},
such techniques typically rely on features that will likely be lost
in the strings of bytes representing binary code after the
compilation process (e.g., variable and function renaming, comments,
and code organization, or the development environment, such as
programming languages and text editors). Identifying the author of a
malware binary might be possible but challenging. Such
identification must be based on features of the malware binary that
are considered to be author specific, which means those features
must show only small variations in the writing of different programs
by the same author and large such variations over the writing by
different authors~\cite{spafford1993software}. That is, authorship
identification requires stylistic features that depend on authorship
of the code, instead of any other properties, such as functionality.
This fact implies that most existing malware analysis techniques
will not be directly applicable to authorship attribution. On the
other hand, several papers show that the stylistic features are
abundant in binaries
\cite{alrabaee2014oba2,caliskan2015coding,rosenblum2011wrote}, and
it may be practically feasible to identify the authorship with
acceptable accuracy. Another challenge unique to malware authorship
attribution is that, while software code may take many forms,
including sources files, object files, binary files, and shell code,
the malign nature of a malware usually dictates the focus on binary
code due to the lack of source code.

In this paper, we investigate the state of the art on binary code
authorship techniques and analyze them. More specifically, we first
present the survey of existing techniques that are related to the
analysis of authorship attribution. This paper covers related work
on different representations of malware, including both source files
and binaries. Second, we also look at a broader range of work on
general purpose malware analysis in order to study which features
are involved. Such a comprehensive study of features will allow us
to consider a rich collection of features before selecting those
which potentially survive the compilation process and are present in
the binary code. Third, we analyze and compare binary authorship
attribution systems
\cite{alrabaee2014oba2,caliskan2015coding,rosenblum2011wrote}.
Besides, we study their applicability to real malware binaries.
Based on our analysis, we provide many important steps that should
be considered by reverse engineers, security analysts, and
researchers when they deal with malware authorship attribution.


\section{Authorship Attribution}~\vspace{-0.55cm}

In this section, we review the state of the art in the broad domain
of authorship attribution, including some techniques proposed for
malware analysis. An important goal of this study is to collect a
rich list of features that are potentially relevant to malware
authorship attribution.


\subsection{Source Code Authorship Attribution}

Investigating source code authorship attribution techniques can help
us understand the features that are likely preserved during the
compilation process. Several studies have shown that certain
programmers or types of programmers usually hold some features of
programming. Examples are layout (spacing, indentation and boarding
characters, etc.), style (variable naming, choice of statements,
comments, etc.) and environment (computer platform, programming
language, compiler, text editor, etc.). The authorship
identification of source codes has been gaining momentum since the
initial empirical work of Krsul~\cite{krsul1997authorship}. Krsul et
al. described different important applications of source code
authorship techniques and found that style-related features could be
extracted from malicious code as well.
Burrows~\cite{burrows2007source} and Frantzeskou et
al.~\cite{frantzeskou2006source} use n-grams with ranking methods.
Burrows and Frantzeskou have both proposed information retrieval
approaches with n-grams for source code authorship attribution.

Kothari et al.~\cite{kothari2007probabilistic} first collected
sample source code of known authors and created profiles by using
metrics extraction and filtering tools. In addition, they used
style-based and character sequences metrics in classifying the
particular developer. Chen et al.~\cite{chen2010author} proposed a
semantic approach for identifying authorship by comparing program
data flows. More specifically, they computed the program
dependencies, program similarities, and query syntactic structure
and data flow of the program. Burrows et
al.~\cite{burrows2009application} introduced an approach named
Source Code Author Profile (SCAP) using byte level n-gram technique.
The author claimed that the approach is language independent and
n-gram profiles would represent a better way than traditional
methods in order to find the unique behavioral characteristics of a
specific source code author. Jang et al.~\cite{jang2011bitshred}
performed experiments to find a set of metrics that can be used to
classify the source code author. They worked on extracting the
programming layout, style, structure, and fuzzy logic metrics to
perform the authorship analysis. Yang et
al.~\cite{yang2010improving} performed experiments to support the
theory that a set of metrics can be utilized to classify the
programmer correctly within the closed environment and for a
specific set of programmers. With the help of programming metrics,
they suggested developing a signature of each programmer within a
closed environment. They used two statistical methods, cluster and
discriminant analysis. They did not expect that metrics gathered for
a programmer would remain an accurate tag for a long time. It is
obvious that a one-time metrics gathering is not enough, as this
should be a continuous task. The practice of authorship analysis
includes metrics extraction, data analysis and classification.

A separate thread of research focuses on plagiarism detection, which
is carried out by identifying the similarities between different
programs. For example, there is a widely-used tool called Moss that
originated from Stanford University for detecting software
plagiarism~\cite{aiken2005moss}. More recently, Caliskan-Islam et
al. ~\cite{caliskan2015anonymizing} investigated methods to
de-anonymize source code authors of C++ using coding style. They
modeled source code authorship attribution as a machine learning
problem using natural language processing techniques to extract the
necessary features. The source code is represented as an abstract
syntax tree, and the properties are driven from this tree.


\subsection{Binary Code Authorship Attribution}

In contrast to source code, binary code has drawn significantly less
attention with respect to authorship attribution. This is mainly due
to the fact that many salient features that may identify an author's
style are lost during the compilation process.
In~\cite{alrabaee2014oba2,caliskan2015coding,rosenblum2011wrote},
the authors show that certain stylistic features can indeed survive
the compilation process and remain intact in binary code, which
leads to the feasibility of authorship attribution for binary code.
The methodology developed by Rosenblum et
al.~\cite{rosenblum2011wrote} is the first attempt to automatically
identify authors of software binaries. The main concept employed by
this method is to extract syntax-based and semantics-based features
using predefined templates, such as idioms (sequences of three
consecutive instructions), n-grams, and graphlets. Machine learning
techniques are then applied to rank these features based on their
relative correlations with authorship. A subsequent approach to
automatically identify the authorship of software binaries is
proposed by Alrabaee et al.~\cite{alrabaee2014oba2}. The main
concept employed by this method is to extract a sequence of
instructions with specific semantics and to construct a graph based
on register manipulation, where a machine learning algorithm is
applied afterwards. A more recent approach to automatically identify
the authorship of software binaries is proposed by Caliskan et
al~\cite{caliskan2015coding}. They extract syntactical features
present in source code from decompiled executable binary. Though
these approaches represent a great effort on authorship attribution,
it should be noted that they were not applied to real malware.
Further, some limitations could be observed including weak accuracy
in the case of multiple authors, being potentially thwarted by light
obfuscation, and their inability to decouple features related to
functionality from those which are related to authors' styles.


\section{Study of Features}

In this section, we present a more elaborated study of features
collected during the literature review.



\subsection{Features of Source Files}

Program source code provides a far richer basis for writer-specific
programming features. Our goal is to determine which features may
survive the compilation process and be helpful for authorship
identification of binary code.

\textbf{Linguistic Features:} Programming languages allow developers
to express constructs and ideas in many ways. Differences in the way
developers express their ideas can be captured in their programming
styles, which in turn can be used for author
identification~\cite{shevertalov2009use}. The linguistic style is
used to analyze the differences in the literary techniques of
authors. Researchers have identified over 1,000 characteristics, or
style markers, such as comments, to analyze literary
works~\cite{can2004change}. Moreover, it has been used to identify
the author by capturing, examining, and comparing style
markers~\cite{holmes1994authorship}.

\textbf{Formatting:} The source code formatting shows a very
personal style. Formatting is also considered as a good way for
programmers to make it easier when reading what was written. These
factors indicate that the formatting style of code should yield
writer-specific features~\cite{krsul1997authorship}: Placement of
statement delimiters, Multiple statements per line, Format of type
declarations, Format of function arguments, and Length of comment
lines.

\textbf{Bugs and Vulnerabilities:} A written program might have
errors or bugs such as buffer overflow, or a pointer to an undefined
memory address. These kinds of issues could be an indicator of the
author.

\textbf{Execution path:} The execution path may indicate the
author's preference in how resolving a particular task through the
selection of algorithms, as well as certain data structures, or
using specific keywords such as {\tt while} or {\tt for}.

\textbf{Abstract Syntax Tree (AST):} AST is an intermediate
representations produced by code parsers of compilers, and thus
forms the basis for the generation of many other code
representations. Such tree forms how statements and expressions are
nested to produce programs. More specifically, it encompasses inner
nodes representing operators (e.g., additions or assignments) and
leaf nodes correspond to operands (e.g., constants or identifiers).

\textbf{Control Flow Graph (CFG):} It describes the order in which
code statements are executed as well as conditions that need to be
met for a particular path of execution to be taken. Statements and
predicates are represented by nodes, which are connected by directed
edges to indicate the transfer of control. For each edge, there is a
label of true, false or unconditional control.

\textbf{Program Dependence Graph (PDG):} It is introduced by
Ferrante et al.~\cite{ferrante1987program}, which has been
originally developed to perform program
slicing~\cite{weiser1981program}. This graph determines all
statements and predicates of a program that affect the value of a
variable at a specified statement.


\subsection{Features of Binary Files}

\textbf{Compiler and System Information:} A unique sequence of
instructions might be an indicator of the compilers. The code may
contain different system calls found only in certain operating
systems. The analysis of binary code may reveal that it was written
in a specific source language such as C++. This can be determined
based on support routines and library calls in the binary code.

\textbf{System Call:} It is considered as programmatic way in which
a computer program requests a service from the kernel of the
operating system it is executed on, for instance, process scheduling
with integral kernel services. Such system calls capture intrinsic
characteristics of the malicious behavior and thus are harder to
evade~\cite{canali2012quantitative}.

\textbf{Errors:} The binary code might have errors or bugs such as
buffer overflow, or a pointer to an undefined memory address. These
kinds of bugs could be an indicator of the author.

\textbf{Idioms:} An idiom is not really a specific feature, but
rather a feature template that captures low-level details of the
sequence underlying a program. Idioms are short sequences of
instructions. A grammar for idiom feature follows the
Backus-Naur~\cite{knuth1964backus} form.

\textbf{Graphlet:} A graphlet is an intermediary representation
between the assembly instructions and the Control Flow Graph, which
represents the details of a program
structure~\cite{prvzulj2004modeling}, and is represented as a small
connected non-isomorphic induced sub-graph of a large
network~\cite{edwards2012historical}. Graphlets were first
introduced by Prvzulj et al.~\cite{prvzulj2004modeling} for
designing two new highly sensitive measures of network locality,
structural similarities: the relative graphlet frequency
distance~\cite{edwards2012historical}, and the graphlet degree
distribution agreement~\cite{rosenblum2011wrote}.

\textbf{n-grams:} The n-gram feature was first used by an IBM
research group (Kephart, 1994). An n-gram is an n-character slice of
a longer string. A string is simply split into substrings of fixed
length $n$. For example, the string 'MALWARE' can be segmented into
the following 4-grams: 'MALW', 'ALWA', 'LWAR', and "WARE".

\textbf{Opcode:} An opcode is the portion of an assembly instruction
that specifies the action to be performed, for instance, {\tt jmp},
{\tt lea}, and {\tt pop}. Opcode sequences have recently been
introduced as an alternative to byte
n-grams~\cite{kruegel2005polymorphic}. Some of the opcodes (e.g.
{\tt push} or {\tt mov}) have a high frequency of appearance within
an executable file. In~\cite{santos2009n} is shown that the opcodes
by themselves were capable to statistically explain the variability
between malware and legitimate software.

\textbf{Strings and Constants:} The type of constants that used in
the literature is integers, which are used in computation, as well
as integers used as pointer offsets. The strings are ANSI
single-byte null-terminated strings~\cite{khoo2013rendezvous}.

\textbf{Register Flow Graph:} This graph captures the flow and
dependencies between the registers that annotated to {\tt cmp}
instruction~\cite{alrabaee2014oba2}. Such graph can capture an
important semantic aspects about the behavior of a program, which
might indicate the author's skills or habits.


\section{Implementation }\label{sec:imp}
This section shows the setup of our experiments and provides an
overview of the collected data.


\subsection{Implementation Environment} The described binary feature extractions are
implemented using separate python scripts for modularity purposes,
which altogether form our analytical system. A subset of the python
scripts in our evaluation system is used in tandem with IDA Pro
disassembler~\cite{FLIRT}. The Neo4j~\cite{neo4j} graph database is
utilized to perform complex graph operations such as \emph{k}-graph
(graphlet) extraction. Gephi~\cite{Gephi} is used for all graph
analysis functions (e.g., page rank) that are not provided by Neo4j.
The PostgreSQL database is used to store extracted features
according to its efficiency and scalability. For the sake of
usability, a graphical user interface in which binaries can be
uploaded and analyzed is implemented.


\subsection{Dataset} The utilized dataset is composed of several files
from different sources, as described below: i) GitHub~\cite{GitHub},
where a considerable amount of real open-source projects are
available; ii) Google Code Jam~\cite{google}, an international
programming competition, where solutions to difficult algorithmic
puzzles are available; and iii) a set of known malware files
representing a mixture of nine different families~\cite{mic}
provided in Microsoft Malware Classification Challenge. According to
existing works, we only examine code written in C/C++. These
programs are either open-source or publicly available, in which case
the identities of the authors are known. Statistics about the
dataset are provided in Table~\ref{tab:s-dataset}.


\begin{table*}[!htbp]\footnotesize \begin{center}\caption {Statistics about the binaries used in the evaluation}\label{tab:s-dataset} \footnotesize
    \begin{tabular}{|l|c|c|c|c|c|} 
    \hline \hline
            \textbf {Source}             & \textbf {\# of authors} &    \textbf {\# of programs}  &  \textbf{\# of functions}   \\
            \hline \hline
                      GitHub                    &   50      &     150     &    40000                \\
    \hline
                      Google Code Jam           &   120     &    550      &    1065      \\
    \hline
                      Malware                   &   9       &     36      &    15000        \\
    \hline
                      Total                     &   179     &    736       &    46065        \\
   \hline \hline

  \end{tabular}
\end{center}
\end{table*}


\subsection{Dataset Compilation} To construct our experimental
datasets, we compile the source code with different compilers and
compilation settings to measure the effects of such variations. We
use GNU Compiler Collection's gcc, Xcode, ICC, as well as Microsoft
Visual Studio (VS) 2010, with different optimization levels.


\subsection{Implementation Phases}

The original binaries are passed to the pre-processing component,
where are disassembled with IDA Pro disassembler. The second
component contains two processes: (1) ASMTODB, which extracts some
specific features (e.g., idioms) from the assembly files, and (2)
BINTODB, which extracts the features directly from the binary files.
The result of this stage is a set of features stored in the
database. This phase also implements the feature ranking which is a
pre-processing phase for classification.


\subsection{Feature Ranking}
Feature ranking is a pre-processing phase for classification. We
assume that there exists a known set of program authors and a set of
programs written by each of them. The task of the feature ranking
algorithm is to associate the identity of the most likely author of
a feature. We extract features from the program assemblies and
binaries as described in the previous section in order to obtain the
feature list associated with a specific author. We apply mutual
information and information gain applied in Rosenblum et
al.~\cite{rosenblum2011wrote} and Islam et
al.~\cite{caliskan2015coding}, respectively.


\subsection{SQL Schema to Store All Features}

Storing, ranking and processing the features in the classification
phase require an appropriate SQL schema. We have chosen the
PostgreSQL database system, and designed our SQL tables, the
relations between them, together with the Features-to-DB APIs, so
that our software modules minimize their interaction with the
database.


\subsection{Authorship Classification}\label{sec : Classification}
The authorship classification technique assumes that a known set of
authors with their program samples are collected. After extracting
and ranking features, as described in the previous subsection, a
classifier is built based on the top-ranked features, producing a
decision function that can assign a label (authorship) to any given
new program based on the given set of known authors.
More specifically, the typical steps for authorship classification
are the following: ~\vspace{-0.3cm}
\begin{enumerate}
 \item Each program is first represented as an integral-valued feature
   vector describing those features that are present in the program.
 \item
 Those features are ordered using the aforementioned ranking algorithm
 based on the mutual information between the features and the known
 author labels. A given number of top-ranked features are selected, and
 others filtered out in order to reduce both the training cost and
 risk of overfitting the data.
 \item A cross-validation is performed on those highly-ranked
   features. Those features would jointly produce a good decision
   function for the authorship classifier.
 \item The LIBLINEAR support vector machine for the actual classification is employed for the actual classification.
\end{enumerate}



\section{Evaluation}  \label{sec:eva}

In this section, we present the evaluation results for the existing
works on binary authorship attribution. Subsequently, we evaluate
the identification and the scalability of existing works. The impact
of evading techniques is then studied. Finally, binary features are
applied to real malware binaries and the results are discussed.

~

\subsection{Accuracy} \label{sec:AAAR}

The main purpose of this experiment is to demonstrate how to
evaluate the accuracy of author identification in binaries.

\textbf{Evaluation Settings.} The evaluation of existing works is
conducted using the datasets described in Section~\ref{sec:imp}. The
data is randomly split into ten sets, where one set is reserved as a
testing set, and the remaining sets are used as training sets to
evaluate the system. The process is then repeated 15 times
(according to existing works). Furthermore, since the application
domain targeted by binary authorship attribution works is much more
sensitive to false positives than false negatives, we employ an
F-measure as follows: \vspace{-0.15cm}
\begin{equation}
F{_{0.5}} = 1.25~.~ \frac{{{P~.~R}}}{0.25P + {R}}
\end{equation}\label{equation7}




\textbf{Existing Works Comparison.} We evaluate and compare the
existing authorship attribution
methods~\cite{alrabaee2014oba2,caliskan2015anonymizing,rosenblum2011wrote}.
For this purpose the source code and the used database are needed.
The source code of the authorship classification techniques
presented by Rosenblum et al.~\cite{rosenblum2011wrote} and
Caliskan-Islam et al.~\cite{caliskan2015anonymizing} are available
at \cite{materials} and \cite{bda}, respectively; however the
datasets are not available. For the system proposed by Alrabaee et
al.(OBA2)~\cite{alrabaee2014oba2}, we have asked the authors to
provide us the source code.

Caliskan-Islam et al. present the largest scale evaluation of binary
authorship attribution in related work, which contains {\it 600}
authors with {\it 8} training programs per author. Rosenblum et al.
present a large-scale evaluation of {\it 190} authors with at least
{\it 8} training programs, while Alrabaee et al. present a small
scale evaluation of 5 authors with {\it 10} programs for each. Since
the datasets used by the aforementioned techniques are not
available, we compare our results with these methods using the same
datasets mentioned in Table~\ref{tab:s-dataset}. The number of
features used in Caliskan-Islam et al, Rosenblum et al, and Alrabaee
et al systems are {\it 4500}, {\it 10000}, and {\it 6500},
respectively.

\begin{figure}[!htbp]
	\includegraphics[scale=0.7]{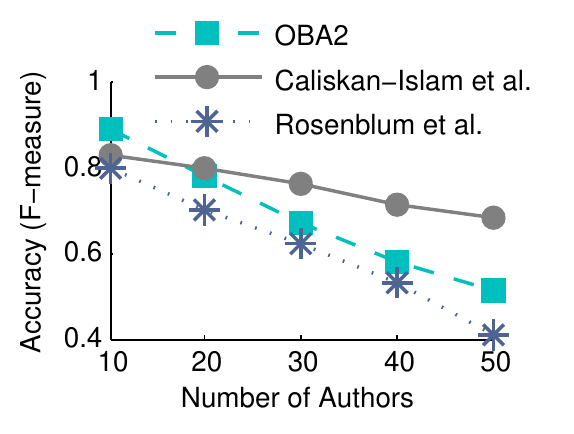}
		\includegraphics[scale=0.7]{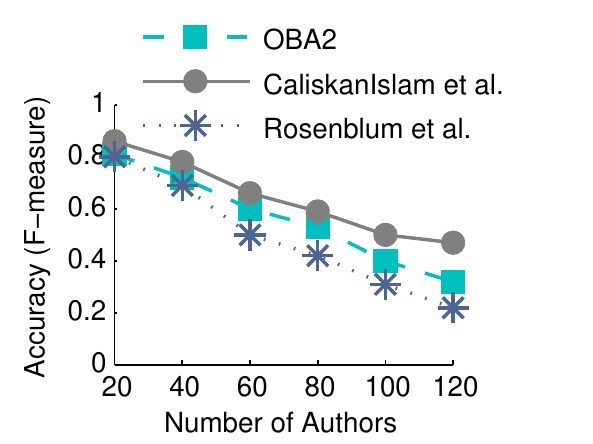}
\includegraphics[scale=0.65]{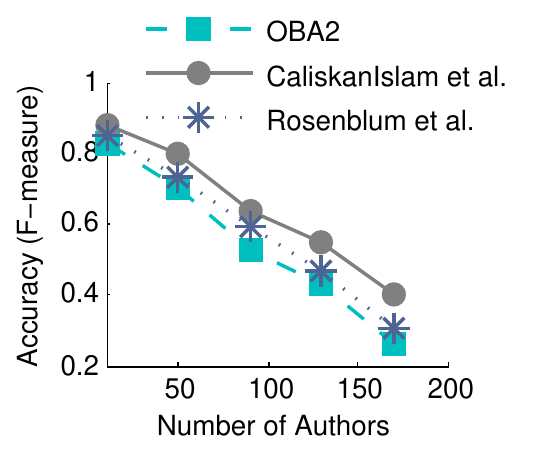}
	\caption{Accuracy results of authorship
attribution obtained by Caliskan-Islam et
al.~\cite{caliskan2015anonymizing}, Rosenblum et
al.~\cite{rosenblum2011wrote}, and OBA2~\cite{alrabaee2014oba2}, on
(a) Github, (b) Google Code Jam, and (c) All
datasets.}\label{fig:results}
\end{figure}

Figure~\ref{fig:results} details the results of comparing the
accuracy between existing methods. It shows the relationship between
the accuracy ($F_{0.5}$) and the number of authors present in all
datasets, where the accuracy decreases as the size of author
population increases. The results show that Caliskan-Islam et al.
approach achieves better accuracy in determining the author of
binaries. Taking all three approaches into consideration, the
highest accuracy of authorship attribution is close to {\it 90\%} on
the Google Code Jam dataset with less than {\it 20} authors, while
the lowest accuracy is {\it 45\%} when {\it 179} authors are
involved.


As can be seen in Figure~\ref{fig:results}, OBA2 achieves good
accuracy when it deals with small scale of authors. For instance,
the accuracy is approximately {\it 84\%} on GitHub dataset when the
number of authors is {\it 30}, while the accuracy drops to {\it
58\%} on the same dataset when the number of authors increases to
{\it 50}. The main reason is due to the fact that the authors of
projects in Github have no restrictions when developing projects.
The lower accuracy obtained by OBA2 is approximately {\it 28\%} on
all datasets when the number of authors is {\it 179}. The accuracy
of Rosenblum et al. approach drops rapidly to {\it 43\%}, whereas
Caliskan-Islam et al. system accuracy remains greater than {\it
60\%}, if the {\it 140} authors on all datasets are considered.


\subsection{Scalability} \label{sec:Scalability}~\vspace{-0.07cm}
We evaluate how well existing works scale up to {\it 1000} authors.
Since in the case of malware, an analyst may be dealing with an
extremely large number of new samples on a daily basis. For this
experiment, we work with {\it 1000} users, of which are authors from
the Google Code Jam. First, we extract the top-ranked features as
described in Rosenblum et al. and Caliskan-Islam et al. approaches,
while the features used by OBA2 are not ranked.

\begin{wrapfigure}{r}{0.5\textwidth}
  \vspace{-20pt}
  \begin{center}
    \includegraphics[scale=0.6]{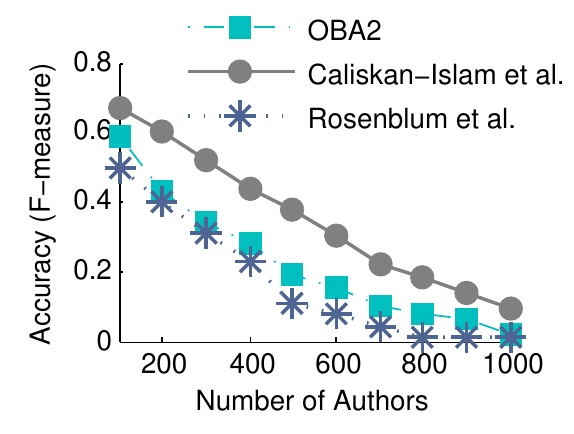}
  \end{center}
  \vspace{-20pt}
  \caption{Large-scale author identification results}\label{fig:large}
  \vspace{-8pt}
\end{wrapfigure}

The results of large-scale author identification are shown in
Figure~\ref{fig:large}. As seen in Figure~\ref{fig:large}, by
increasing the number of authors, all the existing works accuracy
drops significantly. For instance, Rosenblum et al. approach
accuracy drops to approximately {\it 5\%} when the number of authors
is greater than {\it 600} authors. While the accuracy of OBA2
approach drops to {\it 15\%} when the number of authors reaches to
{\it 500}. However, Caliskan-Islam et al. approach accuracy drops to
{\it 20\%} with an increas to over {\it 700} authors.

Through our experiments we have observed that top-ranked features
used in the Rosenblum et al. approach are mixture of compiler and
user features, where leads to higher rate in false positives. OBA2
identifies author according to the way of branch handling.
Therefore, when the number of authors is largely increased,
distinguishing the author based on handling branches becomes limited
and hard. Finally, Caliskan-Islam et al. approach relies on the
features extracted from AST of compiled binary; so with the large
number of authors, these features became common and similar which
make the authorship attribution harder.




\begin{table}[!h]\footnotesize
\centering \caption{Accuracy results before and after applying
refactoring techniques, obfuscation techniques, and different
compilers. \\
(AbET): Accuracy before Evading Techniques, ({$\sim$}): The accuracy
has not affected.} \label{tab:evade}

\begin{tabular}{|l|c|c|c|c|c|c|c|c|c|c|c|c|c|}
\hline \hline \multirow{2}{*}{System} & \multirow{2}{*}{AbET} &
\multicolumn{4}{l|}{\emph{Refactoring}} &
\multicolumn{5}{l|}{\emph{Obfuscation}} & \multicolumn{3}{l|}{\emph{Compiler}} \\
\cline{3-14}
                                        &                   &  RV   &   NM  &  MM   &  All   &  RR   &  IR   &  DCI   &  FCF    & All   &   GCC     &   Xcode    &   ICC    \\
                                        \hline \hline
                \emph{OBA2}                    &     0.81          &  0.62 &  {$\sim$} &  {$\sim$} &  0.62  &  0.64 &  0.74 & {$\sim$}   &  {$\sim$}  &  0.58 &   0.74    &   0.60     &   0.54    \\ \hline
                \emph{Caliskan-Islam}   &     0.79          &  {$\sim$} &  0.72 & 0.71  & 0.70   &  {$\sim$} &  {$\sim$} &  {$\sim$}  &  0.24  &  0.24 &   0.66    &   0.64    &    0.54   \\ \hline
                \emph{Rosenblum}        &     0.66          &  0.60 &  0.58 & 0.55  &  0.4   &  0.62 &  {$\sim$} &  {$\sim$}  &  0.27  &  0.25 &   0.15    &   0.55    &    0.29   \\
                \hline \hline
\end{tabular}
\end{table}


\subsection{Impact of Evading Techniques} \label{sec:evade}
In this subsection, we apply different techniques to evade the
existing systems in order to study their stability. For this
purpose, we randomly choose {\it 50} authors and {\it 8} programmes
for each author. The accuracy results without applying any evading
technique, and with applying evading techniques are shown in Table
\ref{tab:evade}.


 \textbf{Refactoring Techniques}. The adversary may use existing
refactoring techniques to prevent authorship attribution. Hence, we
use chosen dataset for the C$^{++}$ refactoring process
\cite{refact2,refact1}. We consider the techniques of i) Renaming a
Variable (RV), ii) Moving a Method from a superclass to its
subclasses (MM), and iii) extracting a few statements and placing
them into a New Method (NM). Depth explanations of these techniques
are detailed in~\cite{fowler2009refactoring}. We obtain an accuracy
of {\it 81\%} in correctly classifying authors for OBA2 system,
which drops to {\it 62\%} when RV is applied. The reason of this
dropping in accuracy is that variable renaming affects the features
used by OBA2, while OBA2 can tolerate NM, and MM. The accuracy of
Caliskan-Islam et al. approach drops not greatly from {\it 79\%} to
{\it 70\%}. This is due to the fact that their approach decompiles
the binary into source code, and then extracts the features. Hence,
the aforementioned refactoring techniques do not change much in the
abstract syntax tree. However, the approach can tolerate renaming
variables. Finally, Rosenblum et al. approach is the one that is
mostly affected by Refactoring techniques, where the accuracy drops
from {\it 66\%} to {\it 40\%}. Since their approach extracts idioms
from assembly files, any of these techniques will change the idioms
(sequence of assembly instructions) which cause a drop in accuracy.

\textbf{The Impact of Obfuscation.} We are interested in determining
how existing works handle simple binary obfuscation techniques
intended for evading detection, as implemented by tools such as
Obfuscator-LLVM~\cite{junod2015obfuscator}. These obfuscators could
apply Instruction Replacement(IR): replacing instructions by other
semantically equivalent instructions, Dead Code Insertion (DCI),
Register Renaming (RR), spurious control flow insertion, and can
even completely Flatten Control Flow graphs (FCF). Obfuscation
techniques implemented by Obfuscator-LLVM are applied to the samples
prior to classifying the authors. Caliskan-Islam et al. approach is
the most affected approach by FCF technique; since control flow
flattening makes the decompilation process hard, which means the
features cannot be extracted correctly.

\textbf{The Impact of Compilers.} To create experimental datasets
for this purpose, we first compile the source code with GCC, VS,
ICC, and Xcode compilers. Next, the effect of different compilation
options, such as the source of compiler, is measured. The results
show that the approach which is mostly affected by changing the
compiler is Rosenblum et al.'s approach; since this approach does
not distinguish between user functions or compiler functions. For
instance, the accuracy observed through our experiments is 15\%,
when the binaries are compiled with GCC, becuase the GCC compiler
inserts many compiler functions.


\begin{table*}[!h]\footnotesize
\centering \caption{Characteristics of malware datasets. \\
(BF): binary functions, (CF): compiler functions, (LF): library
function. } \label{tab:malware}
\begin{tabular}{|l|l|l|l|l|}
\hline \hline
 Malware              & \# of variants     & \# of BF  & \# of CF   &  \# of LF                       \\ \hline \hline
 {\tt Ramnit}         &   4                &     5285                &    1601                    &    50                           \\ \hline
 {\tt Lollipop}       &   3                &     3510                &    1054                    &    100                             \\ \hline
 {\tt Kelihos}        &   2                &     1924                &    847                     &    74                                      \\ \hline
 {\tt Vundo}          &   4                &     7923                &    2410                    &    219                            \\ \hline
 {\tt Simda}          &   2                &     2100                &    689                     &    105                               \\ \hline
 {\tt Tracur}         &   2                &     1657                &    787                     &    100                               \\ \hline
 {\tt Obfuscator.ACY} &   3                &     2762                &    986                     &   310                               \\ \hline
 {\tt Gatak}          &   2                &     2054                &    860                     &    174                               \\
 \hline \hline
\end{tabular}
\end{table*}


\subsection{Applying Existing Works to Malware Binaries} \label{sec:malware11}
We apply existing works to different sets of real malware: {\tt
Ramnit}, {\tt Lollipop}, {\tt Kelihos}, {\tt Vundo}, {\tt Simda},
{\tt Tracur}, {\tt Obfuscator.ACY}, and {\tt Gatak}. These malware
are selected due to their availability~\cite{mic}. These samples
contain different variants of the same malware so we assume that
these variants are written by the same author or the same group of
authors. Due to the lack of ground truth, we compare outputs of each
approach manually to verify that they belong to same family. Details
about the malware dataset are shown in Table~\ref{tab:malware}. The
number of compiler functions are obtained based on
\cite{rahimian2015bincomp}, while the fifth column shows the number
of library functions acquired by F.L.I.R.T technology~\cite{FLIRT}.
According to Table~\ref{tab:malware}, we can observe that the
percentage of compiler functions is quite high, so a pre-processing
step before applying authorship attribution approaches would be
demanding. For instance, the percentage of compiler functions in
{\tt Lollipop} family is {\it 30\%}. We apply existing works and
cluster functions according to their features by using standard
k-mean. Then we manually analysis the obtained clusters to classify
them to correct/wrong clusters as shown in Table~\ref{tab:1}.


\begin{table}[!h]\footnotesize
\centering \caption{Clustering results based on the features used in
existing systems.\\
(TC): the total number of clusters, (CC): the percentage of correct
clusters, (WC): the percentage of wrong clusters.} \label{tab:1}
\begin{tabular}{|l|l|l|l|l|l|l|l|l|l|}
\hline \hline \multirow{2}{*}{Malware} &
\multicolumn{3}{l|}{\emph{OBA2}} &
\multicolumn{3}{l|}{\emph{Caliskan-Islam}} & \multicolumn{3}{l|}{\emph{Rosenblum}} \\
\cline{2-10}
                                        &  TC      &   CC   &  WC      &  TC   &  CC   &  WC      &   TC     &   CC    &   WC    \\
\hline \hline
                {\tt Ramnit}                & 145  &   60\% &  30\%   &  110  &  47\%  &  50\%    &   208    &   18\%  &   70\%    \\ \hline
                {\tt Lollipop}              & 90   &   75\% &  14\%   &  185  &  59\%  &  38\%    &   220    &   21\%  &   67\%    \\ \hline
                {\tt Kelihos}               & 41   &   88\% &  8\%    &  17   &  90\%  &  4\%     &   75     &   34\%  &   55\%    \\ \hline
                {\tt Vundo}                 & 200  &   62\% & 14\%    &  89   &  28\%  &  68\%    &  384     &   39\%  &   48\%    \\ \hline
                {\tt Simda}                 & 52   &   49\% & 50\%    &  41   &  92\%  &  5\%     &  109     &   42\%  &   51\%    \\ \hline
               {\tt Tracur}                 & 44   &   89\% & 9\%     &  53   &  83\%  &  12\%    &  124     &   51\%  &   40\%    \\ \hline
              {\tt Obfuscator.ACY}          & 30   &   78\% & 21\%    &  45   &  74\%  &  24\%    &   89     &   29\%  &   70\%    \\ \hline
               {\tt Gatak}                  & 29   &   57\% & 34\%    &  51   &  87\%  &  12\%    &   79     &   38\%  &   62\%   \\
                \hline \hline
\end{tabular}
\end{table}


\section{Learnt Lessons and Concluding Remarks}  \label{sec:less}

\textbf{Functionality or styles:}. During the evaluation, we have
observed that the features selected by existing techniques are more
closely related to the functionality of the program rather than the
author's style. This argument may be supported by the evidence that
a basic short program has less features than comparatively bigger,
functionality-oriented programs. This shows that features are
directly related to the size of the program, which usually depicts
functionality but not style
~\cite{alrabaee2014oba2,caliskan2015anonymizing,rosenblum2011wrote}.
In order to avoid this, some existing systems could be used as preprocessing stage~\cite{alrabaee2015sigma,alrabaee2016bingold} applies different steps.

\textbf{Feature pre-processing:} We have encountered top-ranked
features related to the compiler (e.g., stack frame set-up
operation). It is thus necessary to filter irrelevant functions
(e.g., compiler functions) in order to better identify
author-related portions of code ~\cite{rosenblum2011wrote}. To avoid
this, a filtration method based on the FLIRT technology for library
identification as well as a system for compiler functions filtration
such as BinComp~\cite{rahimian2015bincomp} should be used.
Successful distinction between the two groups of functions
(library/compiler and user functions) will lead to considerable
savings in time and will help shift the focus of analysis to more
relevant functions.

\textbf{Application type:} We find that the accuracy of existing
methods~\cite{alrabaee2014oba2,rosenblum2011wrote} depends highly on
the application's domain. For example, in Figure~\ref{fig:results},
superior accuracy is observed for the Google Code Jam dataset where
the accuracy is 77\% in average. This is because the approach used
by Rosenblum et al. extracts SysCalls, which are more useful in the
case of academia/competition code than in other cases. This can be
explained by the authors' choice to systematically rely on external
libraries and to implement, for instance, MFC APIs. The results also
show that Alrabaee et al. rely on the application because their
approach extracts the manner by which the author handles branches;
for instance, the accuracy drops from {\it 82\%} to {\it 57\%} when
Google Code Jam is used. After investigating the source code, we
notice that the number of branches is not big, which makes the
attribution even more difficult.

\textbf{The source of features:} Caliskan et
al.~\cite{caliskan2015anonymizing} use a decompiler to translate the
program into C-like pseudo code via Hex-Ray~\cite{hex}. They pass
the code to a fuzzy parser, thus abstract syntax tree is obtained,
which is the source of feature extraction. In addition to Hex-Ray
limitations~\cite{hex}, the C-like pseudo code is also different
from the original code to the extent that the variables, branches,
and keywords are different. For instance, we find that a function in
the source code consists of the following keywords: ({\tt 1-do,
1-switch, 3-case, 3-break, 2-while, 1-if}) and the number of
variables is 2. Once we check the same function after decompilation,
we find that the function consists of the following keywords: ({\tt
1-do, 1-else/if, 2-goto, 2-while, 4-if}) and the number of variables
is 4. This will evidently lead to misleading features.

\textbf{Misleading Features:} To make things worse, our
re-evaluation results show that many top-ranked features are in fact
completely unrelated to authors' styles. For example, many source
code-level functions do not have their names identified at binary
level, i.e., IDA Pro assigns a name prefixed with "sub" and
postfixed with randomly generated numbers by the compiler.
Experiments show that these functions with random numbers play a
vital role for features to be ranked high by calculating the mutual
information. This discovery shows that this technique may select
many features unrelated to author styles but rather some other
properties, such as compiler-generated functions. 

\textbf{Concluding Remarks:} Binary code authorship attribution is a
less explored problem compared with source code level authorship
attribution due to many facts (e.g., the reverse engineering is time
consuming, having limited features preserved during the compilation
process). In this paper, we have first presented a literature review
relevant to authorship identification of binary and source code.
Subsequently, we introduce the way of extracting binary features.
Then, we deeply analysis and evaluate the existing works on
different scenarios such as scalability. Finally, we applied them to
real set of malware binaries. It is clear that there exist many
features that could potentially be useful to determine malware
authorship. However, the harder part is to verify their
applicability through experimental studies. We must pay special care
to the following issues when we deal with binary authorship
attribution:

\begin{itemize}
  \item Dataset Size: A small amount of training set code might not
    be sufficient to make a good identification and a precise comparison
    unless very unusual indicators are present.

  \item Multiple Authors: The identification of authors in the case of
  multiple authors will be more challenging, since we have to first identify
  code fragments that are written by the same author.

\end{itemize}

\section*{Acknowledgments}\label{ackcon}
The authors thank the anonymous reviewers for their valuable
comments. Any opinions, findings, and conclusions or recommendations
expressed in this material are those of the authors and do not
necessarily reflect the views of the sponsoring organizations.



{\small
 \bibliographystyle{acm}
  \bibliography{references}
}



\end{document}